\documentclass[sigconf]{acmart}
\AtBeginDocument{%
  }
\copyrightyear{2024}
\acmYear{2024}
\setcopyright{rightsretained}

\acmYear{2018}
\acmDOI{XXXXXXX.XXXXXXX}
\acmConference[ICMI '25]{27th International Conference on Multimodal Interaction (ICMI 2025) }{October 13--17, 2025}{Canberra, Australia}
\acmISBN{978-1-4503-XXXX-X/2018/06}

\usepackage{enumitem}

\begin{document}

\title  [fNIRS Analysis of Interaction Techniques in Touchscreen-Based Educational Gaming]{Functional Near-Infrared Spectroscopy (fNIRS) Analysis of \\ Interaction Techniques in Touchscreen-Based Educational Gaming}
\author{Shayla Sharmin}
\email{shayla@udel.edu}
\orcid{0000-0001-5137-1301}
\affiliation{%
  \institution{University of Delaware}
  \streetaddress{South College Avenue}
 \city{Newark}
  \state{Delaware}
  \country{USA}
 }

\author{Elham Bakhshipour}
\email{elhambak@udel.edu}
\orcid{xxx}

\affiliation{%
  \institution{University of Delaware}
  \streetaddress{South College Avenue}
 \city{Newark}
  \state{Delaware}
  \country{USA}
 }
\author{Mohammad Fahim Abrar}

\email{fahim@udel.edu}
\orcid{0009-0009-5157-7807}

\affiliation{%
  \institution{University of Delaware}
  \streetaddress{South College Avenue}
 \city{Newark}
 \state{Delaware}
 \country{USA}
 }

 \author{Behdokht Kiafar}
\email{kiafar@udel.edu}
\orcid{0009-0001-4415-1332}

\affiliation{%
  \institution{University of Delaware}
 \streetaddress{South College Avenue}
  \city{Newark}
  \state{Delaware}
  \country{USA}
 }

\author{Pinar Kullu}
\email{pkullu@udel.edu}
\orcid{0000-0003-3396-8782}

\affiliation{%
  \institution{University of Delaware}
 \streetaddress{South College Avenue}
 \city{Newark}
 \state{Delaware}
 \country{USA}
}
\author{Nancy Getchell}
\email{getchel@udel.edu}
\orcid{xxx}

\affiliation{%
  \institution{University of Delaware}
  \streetaddress{South College Avenue}
  \city{Newark}
  \state{Delaware}
  \country{USA}
 }
\author{Roghayeh Leila Barmaki}

\email{rlb@udel.edu}
\orcid{0000-0002-7570-5270}

\affiliation{%
University of Delaware, Newark, Delaware, USA
 \streetaddress{South College Avenue}
  \city{Newark}
  \state{Delaware}
  \country{USA}
 }

\renewcommand{\shortauthors}{Sharmin et al.}
\begin{abstract}
Educational games enhance learning experiences by integrating touchscreens, making interactions more engaging and intuitive for learners. However, the cognitive impacts of educational gameplay input modalities, such as the hand and stylus technique, are unclear.
We compared the experience of using hands vs. a stylus for touchscreens while playing an educational game by analyzing oxygenated hemoglobin collected by functional Near-Infrared Spectroscopy and self-reported measures.
In addition, we measured the hand vs. the stylus modalities of the task and calculated the relative neural efficiency and relative neural involvement using the mental demand and the quiz score.
Our findings show that the hand condition had a significantly lower neural involvement, yet higher neural efficiency than the stylus condition. This result suggests the requirement of less cognitive effort while using the hand. Additionally, the self-reported measures show significant differences, and the results suggest that hand-based input is more intuitive, less cognitively demanding, and less frustrating. Conversely, the use of a stylus required higher cognitive effort due to the cognitive balance of controlling the pen and answering questions. These findings highlight the importance of designing educational games that allow learners to engage with the system while minimizing cognitive effort. 
\end{abstract}



\begin{teaserfigure}
  \centering
  \includegraphics[width=0.8\textwidth]{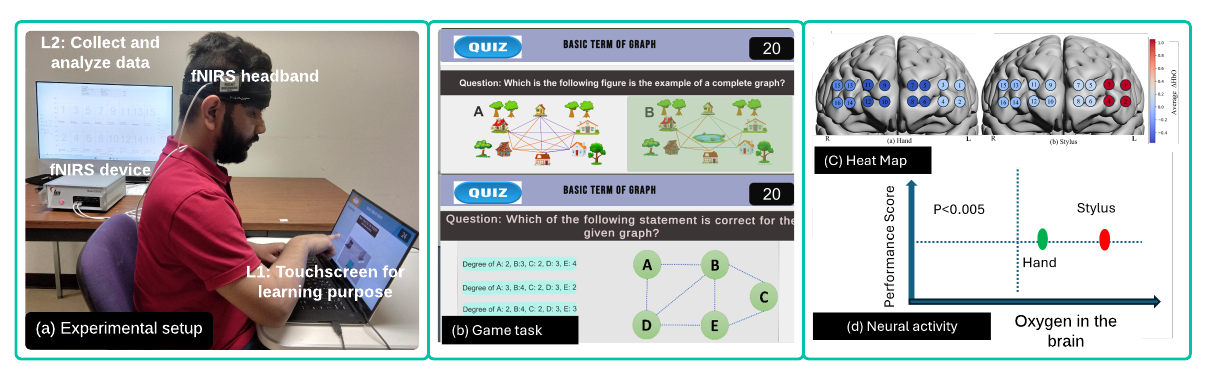}
  \caption{The overview of the study: (a) Experimental setup:  A participant plays an educational quiz game on a touchscreen laptop while functional near-infrared spectroscopy (fNIRS) records prefrontal hemodynamic activity.
(b) Educational Game Environment: Sample game interface with multiple-choice questions.
(c) Hemodynamic Response: Heatmap shows significantly lower oxygen demand when using the hand vs. stylus (d) Cognitive effort: Hand requires significantly less cognitive effort despite similar performance , which indicates neural efficiency differences by input modality.}
  \label{fig:teaser}
  \Description{This teaser figure has four subdivisions showing the study setup, game environment, and findings. (a) Experimental setup: A Participant is playing an educational game on a touchscreen laptop, with a functional near-infrared spectroscopy (fNIRS) device affixed to the forehead. This setup captures hemodynamic responses in the prefrontal cortex (PFC), with data acquisition managed through Cognitive Optical Brain Imaging Studio software on a secondary laptop.
(b) Educational Game Environment: Two quiz questions with options.
(c) Hemodynamic Response: A heat map shows that using the hand required significantly less oxygen flow during the task than using a stylus. (d) Cognitive effort: Hand requires significantly less cognitive effort to get almost same behavioral performance compared to stylus which suggests that input modality affect the neural activities}
\end{teaserfigure}

\maketitle
\section{Introduction}
Touch-based interaction is becoming a more powerful modality due to its resemblance to the traditional pen-and-paper experience. ~\cite{elttayef2016future, 10.1007/978-3-642-33030-8_108, taranta2016dynamic}. 
Incorporating touchscreens into educational games has changed the way we interact while playing. This offers easy touch and movement on the screen, and makes applications more user friendly ~\cite{crescenzi2020emotions}.
Traditional human-computer interaction evaluations, such as questionnaires and think-aloud protocols or user behavior metrics such as reaction times, have limitations, including susceptibility to ambiguity, social pressures, and disruption of the interaction ~\cite{10.1145/2858036.2858525,kosch2023survey,pettersson2018bermuda,yu2023research}. The real-time behavioral metrics such as an functional near-infrared spectroscopy (fNIRS) data offer quantitative insights ~\cite{lloyd2010illuminating,yucel2021best} where the traditional methods fail to accurately represent users' mental states as they are unable to discern between factors such as low concentration or task complexity ~\cite{10.1145/2858036.2858525,fairclough2009fundamentals,frey2013review,pike2012cues} and attention as well as motivation ~\cite{pike2012cues}.

fNIRS, a noninvasive optical imaging technology, measures the changes in concentrations of oxygenated hemoglobin ($\Delta HbO$) and deoxygenated hemoglobin in the prefrontal cortex (PFC) to evaluate hemodynamic responses ~\cite{ lloyd2010illuminating,herff2014mental,yucel2021best, shewokis2015brain, getchell2023understanding,yu2023research, sharmin2024scoping, Shayla10445542}. 
The fNIRS information provides human computer interaction researchers with a powerful tool to explore the complex relationship between cognitive processes and interactive technologies ~\cite{10.1145/2858036.2858525, kosch2023survey, pettersson2018bermuda, sharmin2024scoping}. 
This method assumes that brain or neural activity leads to an increase in oxygenated blood flow, as indicated by increased $\Delta HbO$. This suggests that during task performance, it requires more cognitive effort ~\cite{Shayla10445542, lloyd2010illuminating, herff2014mental, shewokis2015brain, yucel2021best, kosch2023survey, pettersson2018bermuda, yu2023research}.
This connection helps us to understand the brain mechanisms behind cognitive functions, and build technologies and therapies to optimize or monitor them ~\cite{lloyd2010illuminating,yucel2021best}.
The oxygen flow to the PFC can also be used to measure relative neural efficiency, which quantifies cognitive effort and behavioral performance. It can also measure relative neural involvement, which illustrates the relationship among motivation, engagement, and performance ~\cite{paas2003cognitive, shewokis2015brain, getchell2023understanding}. 

This study aimed to investigate the cognitive and usability implications of interaction modalities in touchscreen-based educational gaming. Participants performed quiz tasks in a Unity-developed game environment, using both a hand and a stylus. We used a counterbalanced design within subjects to ensure consistent comparison across the same participant pool. Our contributions are as follows:

   \textbf{Neural Activity Analysis:} The comparison of neural activities associated with hand and stylus input using fNIRS provides insight into their distinct cognitive effort.
   
  \textbf{Cognitive Effort Observation:} The investigation of hand and stylus, influence efficiency and involvement during educational gameplay.
  
    \textbf{Neuro-Usability Link:} Through integrated neural data with usability metrics, an attempt to bridge the gap between cognitive load and user experience in educational tools.

We applied statistical models to oxygen flow, relative neural efficiency, and involvement data to determine whether there were any statistically significant differences between the two groups. 
The research questions we asked are as follows:
\begin{enumerate}[label=$RQ_{\arabic*}$]
\item How does the use of a hand vs. a stylus impact $\Delta HbO$ concentration in the PFC during task performance across different brain regions?
\item How do hand and stylus interactions influence efficiency and involvement during educational game playing?
\item What are the differences in user experience and engagement between hand and stylus?
\end{enumerate}

\section{Related Work} \label{sec:related_work}
The shift from stylus to hand touch on touchscreens is influenced by factors such as precision, user preference, and task requirements. Styluses are ideal for detailed tasks like writing \cite{prattichizzo2015digital}, while hands are more intuitive and easier to use for general interactions \cite{colley2014exploring}. In educational settings, age and engagement often guide this choice \cite{10.1145/2493190.2493222}.

\subsection{Advancements in Touch-based Interaction} \label{sec:advancement_pen_based}
Both pen and hand are widely used in learning contexts, emulating the tactile sensation of pen and paper \cite{aizan2014preschool}.
Modern touchscreen technologies now support intuitive, fast, and direct interactions with minimal cognitive load, while being durable in public and high-usage environments \cite{bhalla2010comparative}.
Advancements such as pressure sensitivity \cite{ahn2019user}, palm rejection \cite{hinckley2014sensing,dempsey2015tactile}, tilt recognition \cite{schwarz2014probabilistic}, and hover detection \cite{pollmann2014hoverzoom} enhance the writing and drawing experience, improve ergonomics, and preserve accessibility.

Integrating pen and hand inputs has enriched user experiences, enabling seamless workflows where users manipulate objects by hand and perform detailed work with a stylus, leveraging the strengths of both modalities \cite{hinckley2010pen+}.
New input methods like Ohmic-touch \cite{ikematsu2018ohmic}, hybrid styluses \cite{campos2022hybrid}, and capacitive enhancements \cite{butler2008sidesight} support these needs.
Designers need to consider interaction context, and display space \cite{forlines2008evaluating}. The ongoing evolution of touch-based technology has increasingly emphasized ergonomics and accessibility, ensuring that a wider range of users, including those with disabilities, can effectively use digital devices \cite{xiong2014ergonomics}. This multi-modal approach to the development of touch technologies underscores their pivotal role in enhancing user interaction in the digital age.

\subsection{Comparison of Interaction Modalities}\label{sec:interaction_related_work}
The choice between using a stylus or direct hand input on touch-sensitive screens is influenced by a number of interrelated factors, including the specific needs of the task, the personal preferences and characteristics of the user, and the use context \cite{colley2014exploring, prattichizzo2015digital, heckeroth2021features}. For tasks involving fine motor control, such as careful writing or drawing, the stylus is typically better suited for accuracy and control \cite{prattichizzo2015digital}. Interestingly, the precision of stylus input is such that digital signatures created in this manner closely match traditional pen-on-paper signatures, with negligible variability in factors like character formation and pen lifts \cite{heckeroth2021features}.

Age significantly influences the choice between stylus and hand touch in educational settings, with younger children preferring styluses for handwriting due to their similarity to pens, while older children choose hand touch for drawing, indicating the need for age-specific tool selection \cite{10.1145/2493190.2493222}. Hand touch is prevalent in modern devices due to their intuitive multi-touch features, such as pinching and tapping; however, comfort and speed vary by finger, emphasizing the need for adaptive interfaces \cite{colley2014exploring}. Integrating multi-touch capabilities into styluses combines precision with ergonomic benefits, although integration challenges persist \cite{song2011grips}. Context-sensitive techniques differentiate stylus and touch inputs for better control, highlighting areas for further research in touch sensitivity and design \cite{hinckley2014sensing}.
Accuracy comparisons show styluses perform better in 2D tasks \cite{Holzinger4588449}, while hand touch is preferred for mid-air interactions among various users \cite{malik2023evaluationaccuracy}. This suggests that stylus vs. hand touch choices should reflect specific contexts and user needs, thereby enhancing the intuitiveness and accessibility of touch interface design.
Buschek et al.'s study on hand and stylus accuracy on mobile touchscreens under various conditions highlighted that styluses, particularly thin ones with hover cursors, were more accurate than hand inputs. Mobility had less impact on accuracy than the input method, slightly reducing the accuracy gap between stylus and hand inputs. Despite styluses' accuracy, many users prefer hand input for its familiarity and ease \cite{10.1145/3161160}.

\subsection{Touch-based Interaction and Education}\label{sec:penbased_education}
Touch-based interaction helps learners understand complex concepts and engage actively, enhancing knowledge and sensory use for a comprehensive learning experience \cite{lee2015does}. This improves both formal and informal education \cite{lee2015does}.
 Touch-based interaction with a stylus or hand on touchscreens provides flexibility and ease, serving as an engaging learning tool \cite{aizan2014preschool}. Touchscreen phones boost education in developing countries due to their affordability, minimal needs, and handwriting support. This highlights the role of touchscreen technology in educational innovation \cite{10.1145/2493190.2493222}.%
Touchscreens provide young children with a simplified and intuitive interface for hands-on learning and direct object manipulation \cite{aizan2014preschool}. Interactive educational apps and games enhance engagement, understanding, and skill development by incorporating gestures such as tapping, dragging, and scaling to boost cognitive and motor abilities \cite{soliman2018evaluating}.
Wang showed that touch-based online games increase learner motivation, knowledge, task completion, and engagement with diverse instructional methods \cite{wang2023effects}. Aizan et al. showed that touch technology boosts learning via enhanced interactivity and motor skills. This offers portability, quick feedback, and extensive digital resources for both individual and group learning \cite{aizan2014preschool}. 
Baham\'{o}ndez et al. found that different age groups prefer to use a stylus or their hand in using touchscreens based on their motor skills for educational tasks \cite{10.1145/2493190.2493222}. This suggests integrating stylus input for writing and drawing in education, enhancing experiences by merging traditional and digital methods. Despite mouse interaction often being more accurate and faster \cite{10.1145/1629826.1629868}, children prefer hands for educational games, using gestures such as tapping, swiping, and dragging \cite{holz2021interaction}. Solaiman and Nathan-Roberts observed no clear preference between stylus and hand for handwriting development in young children \cite{soliman2018evaluating}. Their study showed that hands-on learning could improve letter writing more effectively than using a stylus or pen \cite{soliman2018evaluating}.

\subsection{Importance of Brain Signal Analysis in Interaction Modalities}\label{sec:hemodynamic_related_work}
Human computer interaction research on brain signals requires collaboration between neuroscience and software engineering to decipher brain-system interactions \cite{10.1145/3490554}. Frey et al. assessed mental workload for keyboard vs. touch interfaces using electroencephalogram (EEG) and the NASA Task Load Index (NASA-TLX) questionnaire, finding touch interfaces increased workload and errors across varying game difficulties \cite{10.1145/2858036.2858525}. Alsuradi et al. used a machine-learning classifier on EEG data from a touchscreen device to detect tactile feedback, identifying key channels in the middle frontal cortex and improving classification with time-shifting \cite{9086297}. Alsuradi and Mohamad demonstrated that models trained on EEG data with physical stimuli labels better assess haptic experiences than those with self-reported labels \cite{10113808}. Cho et al. developed a brain-computer interface that distinguishes actual from imagined touch, validating neurohaptics for enhancing virtual and rehabilitative environments \cite{9385331}. Future research is needed to optimize touchscreen technology for educational purposes, including addressing challenges such as unintended touches and tailoring stylus sizes to suit various age groups, environments, and topics.%
\begin{figure} [h]
\centering
  \includegraphics[width=\linewidth]{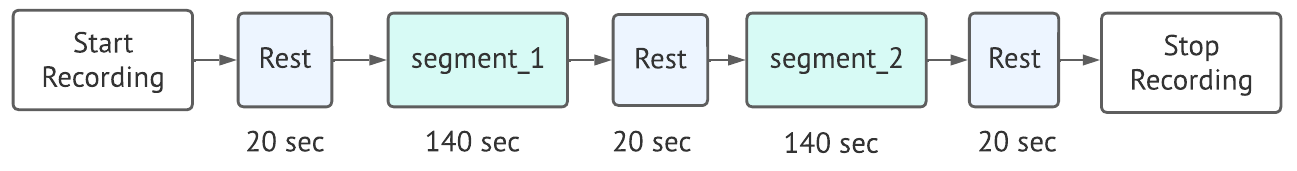}
  \caption{Research protocol for a single session consisting of one quiz. Each quiz is divided into two parts. Each part has a set of four questions, which together last 140 seconds. Rest period is 20-second.  There is no repetition, and the same block structure is used for the second session.}
  \label{fig:Research_Protocol}
  \Description{ In this figure, we have described the research protocol for a single session consisting of one quiz game. Each quiz game is divided into two parts. The figure includes seven boxes to illustrate the pipeline. In the first box, fNIRS data recording begins. This is followed by a 20-second rest period during which participants clear their minds. Then, the first part of the quiz starts, consisting of four questions, each lasting 30 seconds, with a 5-second feedback period, totaling 140 seconds. After this, there is another 20-second rest period for participants to clear their minds. The second part consists of another set of four questions, also totaling 140 seconds, followed by a final 20-second rest period. There is no repetition of questions, and the same block structure is used for the second session.}
\end{figure}
\begin{figure*} [h]

\centering
  \includegraphics[width=\linewidth]{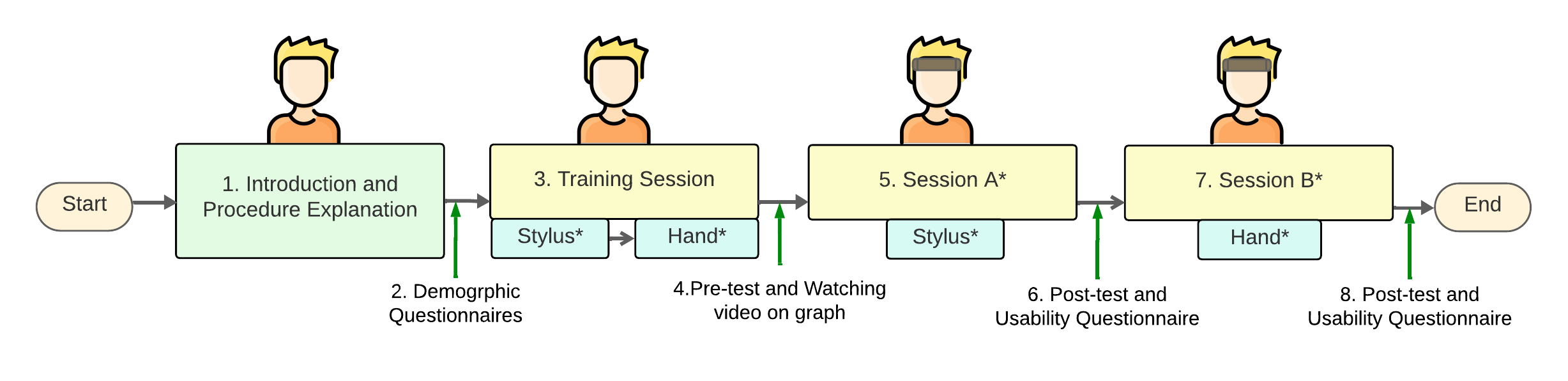}
  \caption{ Overview of the study procedure. The order of the conditions (marked with $^*$) was counterbalanced.}
  \label{fig:Study_Procedure}
  \Description{This figure describes the study procedure. After the study began, the participant was introduced to the study with a detailed explanation and provided their demographic information. Next, they participated in a training session where they used both a stylus and their hand on a touchscreen to understand the study system. Following this, they completed a pre-test and watched a video to learn about the definitions of graph structures. The participant then engaged in a counterbalanced within-subject gameplay, using either their hand or a stylus on a touchscreen laptop. After each session, they completed a post-test and usability questionnaires. The study concluded after two sessions, one using the hand and the other using the stylus.}
\end{figure*}
\section{Materials and Methods} \label{sec:method}
\noindent The experiment was approved by the institutional review board. In the following sections, we describe the participants, apparatus, study procedure, and study design. 

\subsection{Participants}\label{sec:participants}
We used a priori power analysis using \textit{G*Power} to determine the sample size required for interaction effects in repeated measures ANOVA. With an effect size $\eta_{p}^{2}= 0.14$ and a 20\% attrition rate, the target sample size was 16 participants~\cite{faul2007g}.
Participants were recruited through pre-screening based on inclusion criteria, which required them to be healthy adults, proficient in English, and not sensitive to alcohol rub. Informed consent was obtained, and demographic information, including age, education level, gender, and familiarity with graph theory, was collected.
Twenty graduate students initially participated, but two were excluded due to issues with the fNIRS sensor. 

Outliers were identified and removed based on the standard interquartile range (IQR) method ($Q1 - 1.5\times IQR$ or $Q3 + 1.5\times IQR$)\cite{InterquartileRange}. If the average $\Delta HbO$ values of a participant fell outside the range, they were considered outliers.
The final sample comprised 16 graduate students (11 female, 5 male), aged 23–32 years ($M = 27.13 \pm 2.5$). Thus, we removed two more outliers after data collection.  Participants were graduate students aged between 23 and 32 years (11F,$M=27.13 \pm 2.5$). The participants had strong experience using their hands on touchscreens, but only a few were confident using a stylus ($18.75\%$).
More than half had never used games as a learning tool, while others used them occasionally. In terms of graph understanding, only a small number (6.25\%) felt somewhat confident.

\subsection{Apparatus}\label{subsection:Apparatus}\label{sec:apparatus}

In this study, we used two computers and an fNIRS device (Figure \ref{fig:teaser} (a)). We used an 18-channel fNIRS imager 2000S (fNIRS Device LLC, Potomac, MD, USA), four light emitters with wavelengths ($\lambda$) of 730-850 nm and an average power of $<1$ mW, sampled at 2 Hz. The detectors were separated by 2.5 cm, which resulted in a penetration depth of approximately 1.2 cm. The fNIRS sensor band was aligned to match the horizontal and vertical axes of the head with those of the band. Specifically, the sensor's vertical axis was placed at the Fp1 and Fp2 locations delineated in the international 10 - 20 system of cerebral electrode placement.
Data was converted to concentration changes using the modified Beer-Lambert law (mBLL). The educational quiz games were played on a Dell XPS touchscreen laptop (L1) [Intel (R) Core(TM) i7-7700HQ processor at 2.80 GHz]. This laptop supports both hand touch and stylus input with ten touch points. 
We used a desktop computer (L2) [Intel(R) Core(TM) i7-10700T processor at 2.00 GHz] connected to the fNIRS device. Cognitive optical brain imaging studio Software was installed on L2 to gather fNIRS data. fNIRSoft Software (Version 4.9) evaluated the data. 
A Python code was executed on L2 and synchronized with each task to signal the fNIRS device to differentiate between the quiz and rest periods \cite{sharmin2025marker,sharmin2024complexity}.

The Unity environment (version 2019.4.40f1, 64-bit) was employed to develop the educational quizzes. We presented a seven-minute video comprising seven definitions of the basic structure of graphs. 
The participants engaged in two educational quiz games, using their hands in one session and a stylus in the other, with the order of use varying between sessions. During the quiz sessions, the fNIRS data recording started after the participant clicked the ``Start'' button. A 20-second rest period was provided at the beginning as the baseline, during which the participant had to rest and stare at a cross sign on the screen. Subsequently, participants were asked eight questions on graphs. After every four questions, there was a 20-second rest period, displaying a plus (``+'') sign on the screen. Each question lasted thirty seconds, followed by feedback displayed for five seconds. Participants then answered a post-questionnaire before repeating the session with another set of eight conceptually similar questions. 
The game design is illustrated in Figure~\ref{fig:Research_Protocol}.

The quizzes included game elements such as a timer and feedback mechanisms like scores, clapping, confetti, as well as cheering or supporting.

\subsection{Study Procedure}\label{subsection:Study_Procedure}\label{sec:study_prcedure}
Figure~\ref{fig:Study_Procedure} illustrates how this study was conducted. We counterbalanced the session sequence and the interaction module condition (hand and stylus) to mitigate learning curves. The participants signed the consent form, received the introduction, and the explanation of the study in \textbf{Phase 1}. The participants then filled out their demographic questionnaires in \textbf{Phase 2}. Following, in \textbf{Phase 3}, they experienced a demo version where they were given a brief explanation of the two input modalities (hand and stylus) and how to clear their minds during the rest period and focus on the plus (``+'') sign. Additionally, they were instructed on precautions, such as not moving their necks and eyebrows, among other things, on the table. 
In the training session, participants were allowed to practice using the stylus and hand input before starting the quiz. As the task only involved tapping the correct option, this brief training was sufficient for participants to become familiar with the stylus usage.
This allowed the participants to familiarize themselves with the game environment, a basic understanding of the input methods, and objectives, while minimizing learning curve effects during the actual study. 

In the next phase, \textbf{Phase 4}, participants attended a ten-question pre-test on basic terminologies on graphs and did not get answers or feedback, and then watched the recorded video tutorial where they learned the basic terminologies of the graph. 
Subsequently, in \textbf{Phase 5}, the fNIRS headband was placed on the participant's forehead, ensuring a comfortable fit and optimal sensor placement for accurate hemodynamic data collection.
Based on the graph, participants answered the eight quiz questions, four in each part. They interacted with the game using their hands or a stylus. 
After this session, the participants were asked usability questions in a Qualtrics form, and then completed a post-test in \textbf{Phase 6}. In \textbf{phase 7}, they repeat the task with the other set of quiz questions with a similar concept to phase 5, but with the alternate input method. At the end of the study, they answered usability questions in \textbf{Phase 8}. Phases 5 and 6 were counterbalanced. During the quiz, if it was a stylus condition, they held the stylus the entire time.

\subsection{Study Design}\label{subsectiotion:Study_Design}\label{sec:study_design}
The study involved a $2 \times 1$ within-subjects design dependent upon the input method (hand and stylus). In this 
design, each participant experienced all conditions of using both the hand and the stylus across two different sessions of quizzes.
The quiz questions were designed to be conceptually similar, with modified figures. These changes ensured that the cognitive complexity levels and movement required for answering the questions remained similar.
Aside from the training, the amount of entry was $\textrm{16 participants} \times\textrm{2 input methods} \times \textrm{8 questions in each session} = \textrm{256 responses}$.
The independent variable of this study was the interaction modality, which included:
(a) using a hand and 
(b) using a stylus on the touchscreen while playing an educational game. 
Throughout the study, we recorded the participants' quiz scores, task completion times, and hemodynamic responses from the fNIRS as dependent variables.
The Qualtrics survey platform was used to design and distribute the self-reported questionnaires.

\section{Measures \& Instruments}
\subsection{Quantitative Analysis}
For quantitative analysis, we collected hemodynamic responses from the PFC and quiz scores with completion time to observe the neural activities. 

\begin{figure} [H]
\centering
  \includegraphics[width=\linewidth]{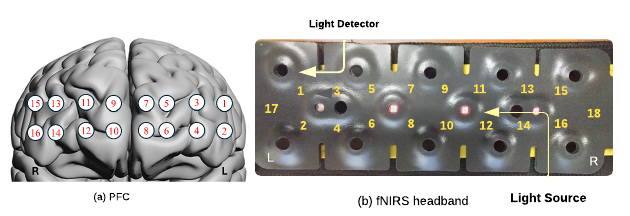}
  \caption{(a) The PFC with four regions: Left Dorsolateral PFC (1-4 channel), Left AnteroMedial PFC (5-8 channel),
Right AnteroMedial PFC (9-12 channel), and Right
Dorsolateral PFC (13-16 channel) (b) 18-channel fNIRS headband sensor pad}
  \label{fig:4channelHeadband}
  \Description{This figure has two subsections. In the left image, (a) we see a frontal view of the brain with 16 circles on the PFC, representing the positions of the 16 channels of the fNIRS headband. In the right image (b), the alignment of all channels of the fNIRS headband is shown, including the light detectors and light sources.}
\end{figure}

\begin{itemize}
\item \textbf{\textit{Behavioral Performance:}} We used Unity to record the start time of each question and the timestamp of the participant’s response. This allowed us to calculate the response time for each question. Unity also logged the participant’s selected answers, which we used to compute performance score accuracy. Using this data, we calculated both the accuracy of the performance scores and the average response time.
\item \textbf{\textit{Hemodynamic Response:}} \label{sec:hemodynamic}
We compared the differences in $\Delta HbO$ concentrations when participants used their hand vs. a stylus across different regions in the PFC using fNIRS. The PFC includes four functional
regions, ~\autoref{fig:4channelHeadband}~(a). The 18-channel headband, showed in ~\autoref{fig:4channelHeadband}~(b), collects oxygen flow in the PFC \cite{getchell2023understanding, koiler2022impact,Shayla10445542}. The four regions are named as Left Dorsolateral PFC (1-4 channel), Left AnteroMedial PFC (5-8 channel),
Right AnteroMedial PFC (9-12 channel), and Right
Dorsolateral PFC (13-16 channel) \cite{getchell2023understanding, Shayla10445542}. Higher oxygen flow suggests more neural activity.
We averaged the $\Delta HbO$ concentrations across channels within each of the four functional regions. Thus, our dataset comprised 128 data points ($\textrm{16 participants}\times \textrm {4 regions} \times 2~\text{input modalities=128}$).

We used a General Linear Model (GLM) to observe the effects of input modality (hand vs. stylus) and PFC on hemodynamic activity ($\Delta HbO$). This framework allowed us to model multiple categorical predictors, estimate effect sizes ($\beta$ coefficients), and directly assess interaction effects between modality and region. Additionally, GLM enabled us to obtain confidence intervals for each predictor. This provides interpretable insights into how different brain regions responded to task conditions.

The GLM was defined by the following equation:
\begin{equation}
    HBO_{ij}=\beta_0+\beta_1* Region_j+\beta_2*(hand/stylus)_i+\epsilon_{ij}
\end{equation}
Where:
\textbf{$HBO_{ij}$ }represented the observed $\Delta HbO$ concentration for the 
$i^{th}$ participant in the $j^{th}$ PFC region.

\textbf{ $\beta_0$} was the intercept, indicating the baseline $\Delta HbO$ concentration for the reference category.

\textbf{ $\beta_1$} captured the effect of each region relative to the reference region.

\textbf{ $\beta_2$ }reflected the effect of using a hand vs. a stylus on $\Delta HbO$ concentration.

\textbf{ $\epsilon_{ij}$} represented the error term, accounting for random variability.

To determine the statistical significance of the hand vs. stylus comparison, we used a t-test where the p-value was associated with the $\beta_2$ coefficient.
For the analysis, we fitted our data using an Ordinary Least Squares (OLS) model using the \textit{smf.ols()} function and used \textit{statsmodels} for statistical analysis in Python to analyze the effect of input modality on $\Delta HbO$ concentration across four channels \cite{statsmodels2025}. 
The coefficients ($\beta_0$, $\beta_1$, and $\beta_2$) were estimated using the OLS method, which minimized the sum of squared residuals. 
We used ``Identity'' as our link function \cite{salinas2023generalized}.
The identity link function models the expected value of the dependent variable ($\Delta HbO$) directly as a linear combination of predictors
without transformation. 

\item \textbf{\textit{Relative Neural Efficiency (RNE) and Relative Neural Involvement (RNI):}}

Cognitive effort can be explained by using two metrics named Relative Neural Efficiency (RNE) and Relative Neural Involvement (RNI) with the behavioral performance~\cite{getchell2023understanding, koiler2022impact, reddy2022individual}. Performing a task efficiently with lower cognitive effort indicates high RNE, where lower RNE suggests high effort and low performance. On the other hand, high RNI indicates the user is more engaged during a task~\cite{paas2003cognitive, shewokis2015brain,getchell2023understanding,koiler2022impact}. Thus, understanding RNE and RNI may help to optimize learning strategies, improve engagement, and enhance cognitive performance.

To find RNE and RNI, we averaged $\Delta HbO$ to measure oxygen demand during the task. This helped us to see how much oxygenated hemoglobin was in the blood in the overall PFC during the learning period. 
Each question had a weight of 50 marks and a time limit of 30 seconds. 

These values were converted to Z-scores to standardize the data. Our aim was better performance with less cognitive effort.  We used the mean task performance score to calculate the standardized performance index ($P_z$) as shown in~\autoref{eq:p_z}. We also computed the inverse of the mean $\Delta HbO$ values to reflect reduced effort, and used this in calculating the standardized average oxygenated hemoglobin index ($M_z$), as shown in~\autoref{eq:ce_z} \cite{AKSOY2025103486,paas2003cognitive, shewokis2015brain,getchell2023understanding,koiler2022impact}. In the X-axis, we plot the standardized average $\Delta HbO$ and on the Y-axis, we plot the standardized performance level $P_z$.
Then, we used a Euclidean transformation from vector geometry.
The equation for RNE in~\autoref{eq:rne} and RNI in~\autoref{eq:rni} represent projections of the vector onto two rotated coordinate axes: one axis captures efficiency (RNE line, running bottom-left to top-right) and the other captures involvement (RNI line, bottom-right to top-left). These rotated axes help to separate two cognitive dimensions: how effectively someone learns, and how deeply involved they are.

\begin{flalign}
&\hspace{2cm} P_z = \dfrac{Score_{i} - Score_{\text{GM}}}{Score_{\text{SD}}} \hspace{1cm} & \label{eq:p_z} \\
&\hspace{2cm} M_z = \dfrac{\dfrac{1}{\Delta HbO_{i}} - \dfrac{1}{\Delta HbO_{\text{GM}}}}{\dfrac{1}{\Delta HbO_{\text{SD}}}} \hspace{1cm} & \label{eq:ce_z} \\
&\hspace{2cm} RNE = \dfrac{P_z - M_z}{\sqrt{2}} \hspace{1cm} & \label{eq:rne}\\
&\hspace{2cm} RNI = \dfrac{P_z + M_z}{\sqrt{2}} \hspace{1cm} & \label{eq:rni}
\end{flalign}
\end{itemize}

\subsection{Self-reported Measures \& Instruments}
We used the System Usability Scale (SUS) \cite{brooke1996sus}, ten questions with a 5-point Likert-scale ranging from `strongly disagree' to `strongly agree.' On a scale of 0 to 100, the final SUS score was determined (0–50 = not acceptable, 51–67 = poor, 68 = okay, 69–80 = good, 81–100 = exceptional) \cite{brooke1996sus}. We also used the NASA-TLX, which consists of six subscales: mental demand, physical demand, temporal demand, performance, effort, and frustration, to assess the mental workload of participants during learning procedures. This assessment has six questions to produce an overall workload rating of 0-100 \cite{hart1988development}.

\section{Results \& Findings}
\subsection{Quantitative Analysis}
\subsubsection{\textbf{Behavioral Performance}}

\textit{Performance Score Accuracy:}
Participants showed slightly higher accuracy with hand input $(M = 0.688, SD = 0.465)$ than with stylus input $(M = 0.625, SD = 0.486)$. However, this difference was not statistically significant ($t(15) = 1.03, p = .304$).

\textit{Response Time:}
Participants also responded slightly faster with stylus input $(M = 6.67 seconds, SD = 8.19)$ than with hand input $(M = 7.19 seconds, SD = 8.69)$. However, the difference was not statistically significant.
These results suggest that, behaviorally, participants performed comparably regardless of the input modality used.

\subsubsection{\textbf{Hemodynamic Response}}

\begin{figure} [h]
\centering
  \includegraphics[width=0.9\linewidth]{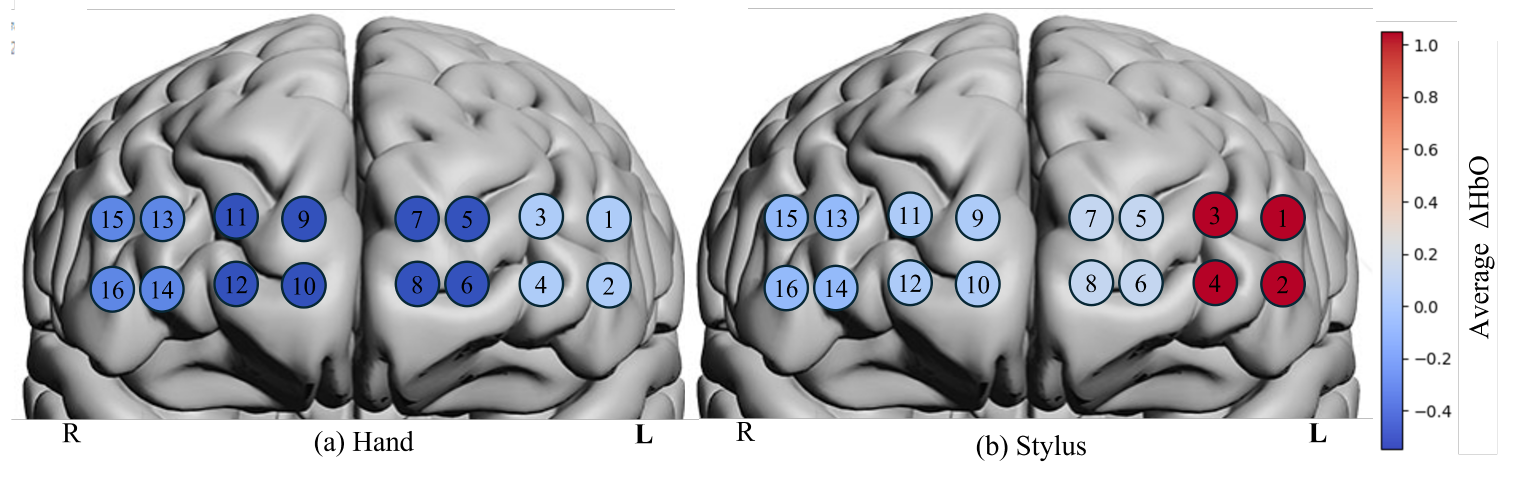}
  \caption{Heatmap of $\Delta HbO$ across four regions of the PFC: (a) hand condition and (b) stylus condition. The stylus condition exhibited overall significantly higher $\Delta HbO$ which signifies greater neural activity compared to the hand condition ($p < 0.05$) across the PFC.}
 \Description{This figure shows two brain images, each with 16 channels across four regions as indicated by the fNIRS headband placement. Each map displays various numbered regions with a heatmap color scale ranging from blue to red, indicating the level of hemodynamic response, with blue representing lower responses and red representing higher responses. In the left image (a), the heatmap represents the hand condition, while in the right image (b), the heatmap represents the stylus condition. Comparing the two figures, we observe that the stylus heatmap (right) shows more neural activity than the hand condition heatmap (left). In both conditions, the Left Dorsolateral (LDL) region shows higher $\Delta HbO$ levels compared to other regions, with the LDL in the stylus condition having the highest $\Delta HbO$ value.}
  \label{fig:heatMap}

\end{figure}

The GLM analysis indicated that task modality (hand vs. stylus) had significant effects on $\Delta HbO$ across the PFC. Participants showed more significant changes in $\Delta HbO$ when using the stylus compared to the hand ($\beta_2 = 0.16, t(123)= 2.08, p = 0.04$, 95\% CI [0.008, 0.312]). 
  The coefficient $\beta_2 (0.16)$ was consistently positive, which suggests higher cognitive effort with stylus use across PFC. 
\autoref{fig:heatMap} shows brain activity maps for using a hand vs. a stylus. Each map displays various numbered regions with a heatmap color scale ranging from blue to red, indicating the level of hemodynamic response. Blue represents lower responses, while red represents higher responses. In Figure ~\ref{fig:heatMap}(a), most brain regions show a lower response in the hand condition. This result suggests less cognitive effort. Figure ~\ref{fig:heatMap}(b) indicates that using the stylus leads to significantly higher cognitive effort ($M=0.14 \pm 0.4$) than a hand ($M=-0.02 \pm 0.3$). 

\subsubsection{\textbf{Relative Neural Efficiency (RNE) and Relative Neural Involvement (RNI)}} 

\begin{figure}[h]
   \centering  \includegraphics[width=0.9\linewidth]{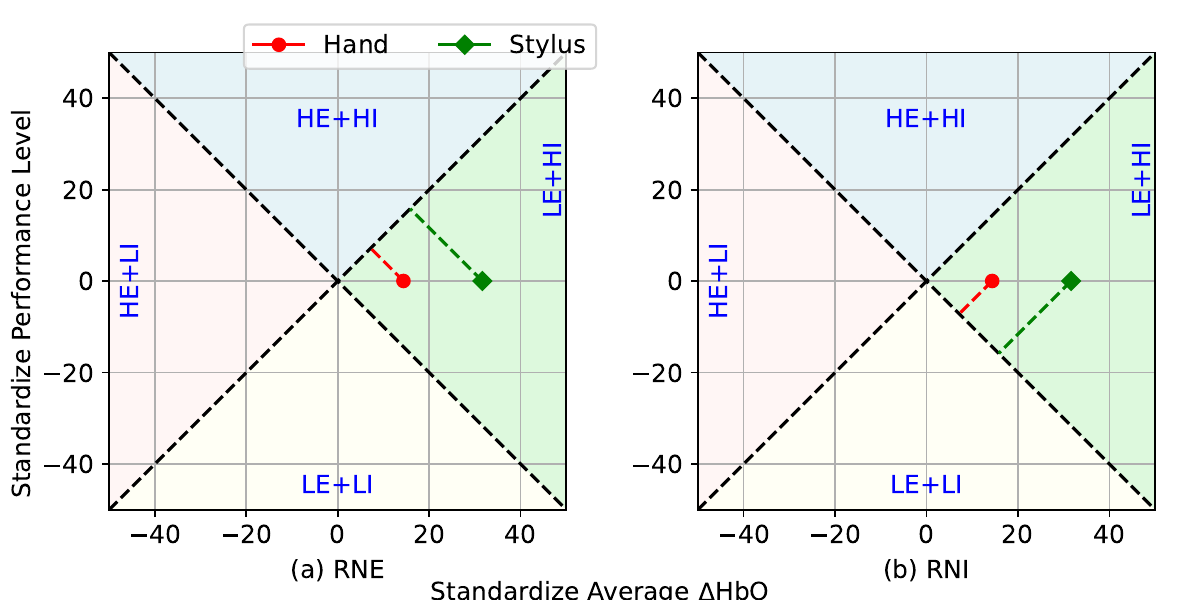}       
    \caption{Relative neural efficiency and involvement: (a) relative neural efficiency (RNE), where the hand condition shows significantly higher neural efficiency ($p < 0.05$) with lower cognitive effort, and (b)  lower Relative Neural Involvement (RNI) compared to the stylus condition ($p < 0.05$).  The red dot represents the hand, and the green diamond represents the stylus condition. The cognitive effort is higher in the stylus although the performance level is almost the same.}

   \Description{The figure consists of three subplots comparing Relative Neural Efficiency (RNE) and Relative Neural Involvement (RNI) between Hand and Stylus conditions. Subplot (a) shows the RNE graph with normalized cognitive effort on the x-axis and normalized performance level on the y-axis. The Hand condition (red dot) is positioned closer to the origin, indicating lower cognitive effort compared to the Stylus condition (green dot), which is farther along the cognitive effort axis, indicating higher effort. Subplot (b) displays the RNI graph, where the Stylus condition (green dot) requires higher neural involvement than the Hand condition (red dot), which is closer to the lower left quadrant. }
    \label{fig:rne_rni}
\end{figure}

We analyzed RNE and RNI by calculating the distance of each participant's data point from the neutral diagonal ($y = x$) in the performance–cognitive effort space. This method captures the tradeoff between cognitive effort and task performance ~\cite{shewokis2015brain}.
 This approach focuses on the spatial relationship between individual participant points and the diagonal axis with a precise view of efficiency and involvement.
Following a Shapiro-Wilk test, the RNI and RNE data conclusively were not normally distributed. Thus, a non-parametric Mann-Whitney U-test was conducted. Participants showed significantly higher RNE during the hand condition compared to the stylus $(U = 18.0, p < .001)$. RNI was also significantly lower during hand use $ (U = 229.0, p < .001)$, which suggests that the stylus required greater neural involvement to achieve similar task performance. 
~\autoref{fig:rne_rni} displays RNE and RNI on Cartesian planes, where distance from the origin represents the overall neural efficiency or involvement. A shorter distance represents an efficient cognitive state \cite{koiler2022impact, reddy2022individual, shewokis2015brain}.

The Cartesian plot in \autoref{fig:rne_rni} (a) (RNE) and (b) (RNI) is divided into four parts: high efficiency and low involvement (HE+LI);  low efficiency and low involvement (LE+LI);  high efficiency and high involvement (HE+HI); low efficiency and high involvement (LE+HI).
The result shows that the performance level is almost the same in both conditions.
 From \autoref{fig:rne_rni}(a), we can see that despite similar performance levels, the hand requires less oxygen. Which means this requires less cognitive effort and thus more RNE compared to the stylus. This can be shown visually by its proximity to the origin on the cognitive effort axis. Conversely, the stylus, positioned further from the origin, indicates that it requires more effort. Additionally, the hand condition shows significantly lower involvement in \autoref{fig:rne_rni}(b) compared to the hand condition. This result suggests that it engages more cognitive resources overall.

\subsection{Self-reported Measures \& Instruments}
We performed paired sample t-tests on the self-reported questionnaire data using Python. This statistical test compares the means between two groups (hand and stylus) for usability and task load.
\begin{figure*}[ht]
    \centering
    \includegraphics[width=0.6\textwidth]{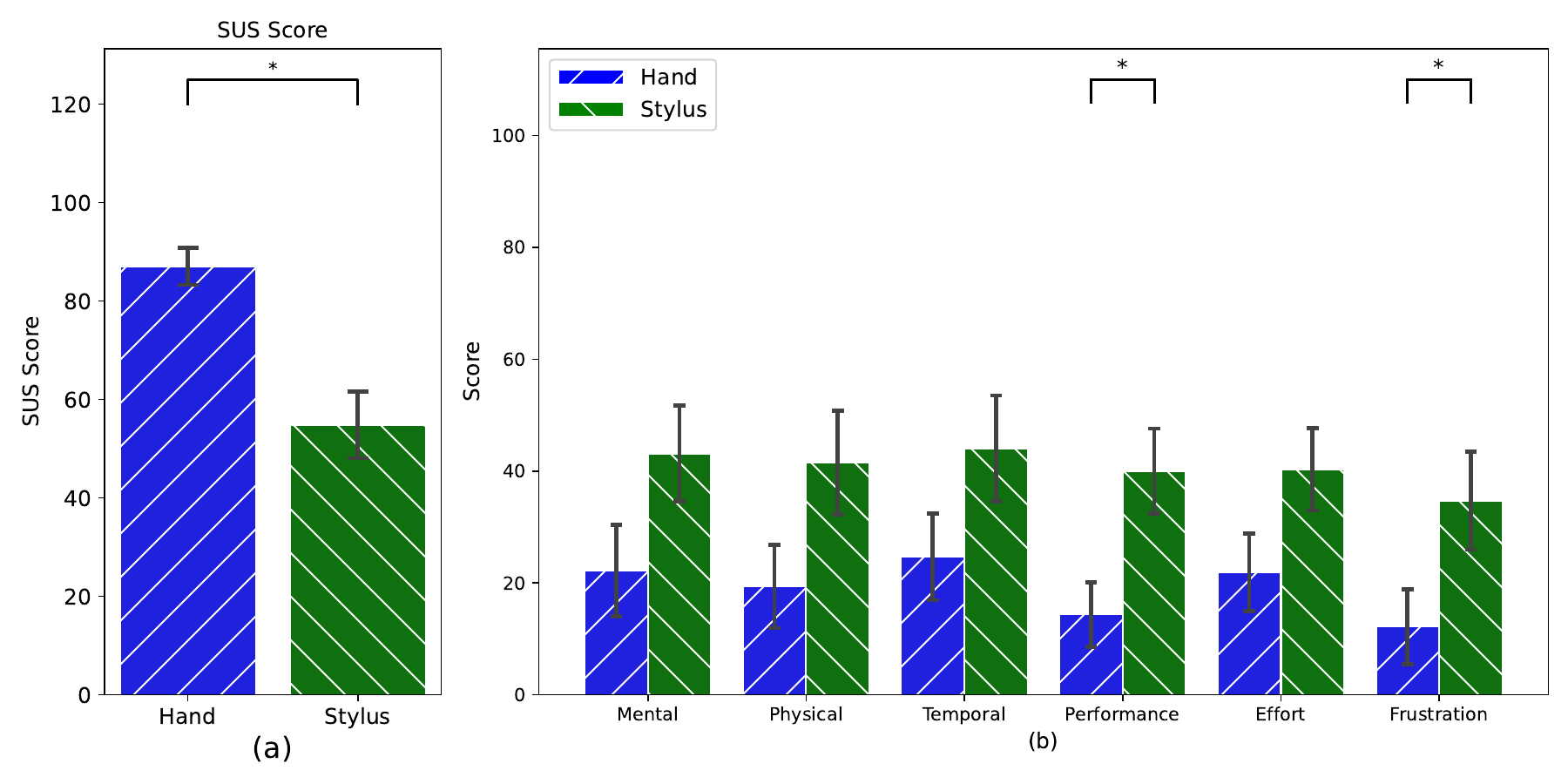}
   \caption{Questionnaire results: (a) System Usability Score (SUS): higher SUS scores indicate better usability and (b) NASA-TLX: lower scores show a reduced task load in accomplishing the task.}
       \label{fig:sus_nasa}
    \Description{This figure shows two bar graphs comparing the Hand and Stylus inputs. In graph (a), the System Usability Score (SUS) is displayed, with higher scores indicating better usability. The Hand input has a significantly higher SUS score than the Stylus, suggesting it is more usable. In graph (b), the NASA Task Load Index (NASA-TLX) scores are presented, evaluating six factors: Mental, Physical, Temporal, Performance, Effort, and Frustration. The Hand input generally shows lower scores across all factors, indicating a reduced task load compared to the Stylus input. The results highlight significant differences in the Performance metric.}
\end{figure*}

\textbf{Usability:} The usability assessment using the SUS showed that the hand condition had a significantly higher average usability score ($M = 87.03 \pm 15$) than the stylus ($M = 54.8 \pm 27.2$) at $p = 0.0003$. See \autoref{fig:sus_nasa}~(a). 
\textbf{Task Load:}
The overall workload scores from NASA-TLX showed a lower average value
($M = 19.22 \pm 4.76$) for the hand than the stylus ($M = 40.47 \pm 3.64$) (lower scores indicate lower workload). The results show that the stylus condition exhibits higher average scores for all six metrics (mental, physical, temporal, performance, effort, and frustration (\autoref{fig:sus_nasa} (b)). 
In terms of performance and frustration, the result showed a statistically significant difference at $p<0.05$. The result suggests that users performed better with a hand ($M = 14.37 \pm 5.77$) compared to a stylus ($M = 40.0 \pm 7.61$). Additionally, participants reported significantly higher frustration rates for the stylus ($M = 34.69 \pm 8.72$) than for their hand ($M = 12.0 \pm 6.68$).

\section{Discussion}

We investigated how input modalities (hand vs. stylus) impact neural activity, cognitive effort, and user experience in an educational game. Our findings show no significant difference in performance between modalities, suggesting both are equally effective for simple tap-based quiz tasks.

\textbf{$RQ_1$} focused on the hemodynamic response in educational gaming with two input methods (hand and stylus). 
Higher $\Delta HbO$ suggests higher oxygen flow to the brain, which is the result of higher neural activity \cite{kerr2022cognitive, Shayla10445542}.
In our study, the change in blood oxygenation levels ($\Delta HbO$) showed a significant decrement with hand usage compared to the stylus. 
The higher $\Delta HbO$ while using a stylus suggests that handling a stylus can require more precise motor control, fine motor skills, and coordination than using the hand directly. Thus, it is observed that the type of input method may change the cognitive effort in educational game studies.
According to the demographic data, most participants were not regular stylus users. While this unfamiliarity could have potentially influenced their performance by increasing cognitive effort, our data did not reflect any significant disruption. Both response time and performance scores did not show any significant differences. This indicates that prior exposure to the stylus did not negatively affect their task performance, particularly for simple actions like tapping. However, this also highlights a critical design consideration: learning materials and interactive tools should be designed in such a way that users invest their effort primarily in the learning process, not in struggling to operate the interface itself.

\textbf{$RQ_2$} focused on observing cognitive effort by measuring relative neural efficiency (RNE) and relative neural involvement (RNI). The results suggest that using the hand required less cognitive effort, as reflected in higher RNE and lower RNI values. On the other hand, the stylus condition showed higher neural involvement, but the performance score was almost the same as the hand condition. In the stylus condition, participants had to manage both holding the stylus and responding to quiz questions simultaneously, which required more tasks and led to more cognitive effort \cite{schwarz2014probabilistic, koiler2022impact, getchell2023understanding}. This suggests that hands are more natural or intuitive in educational game play and also highlights the importance of interaction methods that support efficient performance with lower cognitive effort.

\textbf{$RQ_3$} focused on the impact of self-reported usability and task load measures of different input methods in playing educational games. The SUS score showed that users found the hand-based input more intuitive and usable, with a good usability score. The NASA-TLX results indicated that the hand condition required less effort and caused less frustration. Together, these findings suggest that hand-based input is more efficient and less demanding cognitively than stylus-based interaction for educational gameplay.

According to the results of \textbf{$RQ_1$}, \textbf{$RQ_2$}, and \textbf{$RQ_3$}, the hand condition provides better outcomes than the stylus in both objective and subjective evaluations. Lower $\Delta HbO$ levels with hand usage are revealed by hemodynamic data ($RQ_1$), which suggests less cognitive load. This is further supported by cognitive effort measurements ($RQ_2$). These results suggest that the natural and intuitive usage of the hand leads to higher efficiency (RNE) and lower involvement (RNI) than stylus usage. These findings are supported by self-reported measures ($RQ_3$), as NASA-TLX results show less task load and annoyance, and SUS ratings show improved usability for hand input.
According to these findings, it might be easier to use the hand as an input method in educational games. The degree to which subjective impressions and objective facts agree demonstrates the importance of taking both viewpoints into account when assessing interaction techniques. This highlights the possibilities of hand-based input in learning games and provides information for developing systems that try to reduce cognitive effort while increasing effectiveness and user satisfaction.


Our findings have broader implications for educational technology and natural user interface design. The lower cognitive effort in the hand condition shows that familiarity and intuitiveness are important for user experience and performance. These results contribute to human-computer interaction by showing that neuronal efficiency improves when input complexity is reduced. The study also shows that neurophysiological data, like fNIRS, can validate design choices that are often based on subjective input. By showing cognitive load differences between hand and stylus, our work supports a user-centered approach to input selection. This neuro-informed view can guide future studies on input methods like mouse and interface designs, and help develop adaptive educational tools that match interaction styles to users’ cognitive and behavioral performance.

\textbf{\textit{Limitations and Future Work:}}

This study may have been limited by its small sample size. Differences in participants’ age, education, and background may have influenced the results. Future studies will include a more diverse and balanced group. A laptop setup may have reduced natural interaction; using a tablet or iPad may create a more intuitive experience. Future work should test these devices and explore real-world educational tasks such as drawing, dragging, or multitasking. Other physiological signals, such as eye gaze or facial expressions, should also be explored. Combining these with $\Delta HbO$ and user experience scores may provide deeper insights. This study focused only on hand and stylus input. We did not include mouse, keyboard, or virtual reality, which are common in learning environments. Their exclusion limits the generalizability of our results.


\section{Conclusion}
In conclusion, this study evaluated hand and stylus usage in an in-house educational game, measuring oxygenated hemoglobin concentration ($\Delta HbO$), relative neural efficiency, and involvement alongside feedback from user experience measures. The findings indicate a statistically significant difference in $\Delta HbO$, relative neural efficiency and involvement between the two input methods. A significant difference was shown in user experience metrics, as evidenced by SUS and NASA-TLX scores. In particular, relative neural efficiency suggests that the stylus requires more motor skill and coordination, potentially increasing neural involvement through higher oxygen flow compared to using the hand, which is more natural and intuitive. 
These findings highlight the importance of accounting for cognitive and physiological factors in educational games and natural user interfaces. Understanding how input modalities influence user performance and experience can help developers optimize learning outcomes. This study provides valuable insights for designers seeking to choose input modalities that enhance user performance and learning in educational games. 

\textbf{Safe and Responsible Innovation Statement}
The study was approved by the IRB and conducted with strict privacy safeguards. We used non-invasive fNIRS to examine how input modalities affect cognitive effort in educational gameplay. Data collection followed ethical protocols, with attention to participant diversity. We promote transparency and responsible use of neural data in educational context.

\section*{Acknowledgment}
We express our gratitude to the study participants and lab members. We also thank the National Science Foundation for its support (\#$2222661-2222663$,   \#$2321274$, and \#$2426003$). 
 Any opinions, findings, and conclusions expressed in this material are those of the authors and do not reflect the views of sponsors.



\balance

  \bibliographystyle{ACM-Reference-Format}
  \bibliography{bibfile}

\appendix

\end{document}